\documentclass{a3l}

\newtheorem{problem}{Problem}

\newcommand{\K}{\mathbb{K}}
\newcommand{\F}{\mathbb{F}}
\newcommand{\x}{\mathbf{x}}
\newcommand{\z}{\mathbf{z}}

\newcommand{\E}{\mathbb{E}}
\newcommand{\D}{\mathbb{D}}

\newcommand{\Ker}{\operatorname{Ker}}

\begin{document}
%
\title{Computation of unirational fields}
\author[1]{Jaime Gutierrez}
\author[1]{David Sevilla}
\address[1]{Faculty  of Science, University of Cantabria, E-39071 Santander, Spain}
%
\shorttitle{Computation of unirational fields}

\shortauthor{J. Gutierrez and D. Sevilla}
\maketitle

\begin{abstract}
In this paper we present an algorithm for computing all algebraic
intermediate subfields in a separably generated unirational field
extension (which in particular includes the zero characteristic
case). One of the main tools is Gr\"obner bases theory, see
\cite{BW93}. Our algorithm also requires computing computing
primitive elements and factoring over algebraic extensions.
Moreover, the method can be extended to finitely generated
$\K$-algebras.
\end{abstract}

\section{Introduction}

The goal of this paper is to study the problem of computing
intermediate fields between a rational function field and a given
subfield of it. This computational problem has many applications,
not only in other areas of mathematics like Algebraic Geometry, but
also in Computer Aided Geometric Design. The question of the
structure of the lattice of such intermediate fields is of
theoretical interest by itself; we will focus on the computational
aspects, like deciding if there are proper intermediate fields and
computing them in the affirmative case.

In the univariate case, the problem can be stated as follows: given
$f_1,\ldots,f_m\in\K(t)$, find a field $\F$ such that
$\K(f_1,\ldots,f_m)\varsubsetneq\F\varsubsetneq\K(t)$. By L\"uroth's
Theorem this is equivalent to the problem of decomposing the
rational functions. Algorithms for decomposition of univariate
rational functions can be found in \cite{Zip91} and \cite{AGR95}.

In the multivariate case, the problem can be stated as:

\begin{problem}
Let $\K$ be a field and $\K(x_1,\ldots,x_n)=\K(\x)$ be the rational
function field in the variables $\x=(x_1,\ldots,x_n)$. Given
rational functions $f_1,\ldots,f_m\in\K(\x)$, compute a proper
unirational field $\F$ between $\K(f_1,\ldots,f_m)$ and $\K(\x)$, if
it exists.
\end{problem}

Any unirational field is finitely generated over $\K$ (see
\cite{Nag93}). Thus, by computing an intermediate field we mean that
such a finite set of generators is to be calculated. Regarding
algorithms for this problem, see \cite{MS99}, where the authors
generalize the method of \cite{AGR95} to several variables, by
converting this problem into the calculation of a primary ideal
decomposition. Primary ideal decomposition can be computed by
Gr\"obner Bases. The book \cite{BW93} by T. Becker and Volker
Weispfenning is an excellent reference guide to this important
theory  and their application.

It is not difficult to realize that the solution of the problem is
trivial and uninteresting for most choices of $f_1,\ldots,f_m$,
since it is easy to construct infinitely many intermediate fields
when the transcendence degree of $\K(f_1,\ldots,f_m)$ over $\K$ is
smaller than $n$. Due to this, we will focus on the following
version of the problem.

\begin{problem}\label{prob-alg}
Given functions $f_1,\ldots,f_m \in\K(\x)$, find all the fields $\F$
between $\K(f_1,\ldots,f_m)$ and $\K(\x)$ that are algebraic over
$\K(f_1,\ldots,f_m)$.
\end{problem}

There are finitely many algebraic intermediate fields if the
original extension is separable.

The special case of Problem \ref{prob-alg} when the transcendence
degree of $\K(f_1,\ldots,f_m)/\K$ is 1 has been treated in
\cite{GRS01}. In this case a generalization of L\"uroth's Theorem
applies so the problem is equivalent to the so-called
uni-multivariate decomposition. The paper \cite{GRS02} provides a
very efficient constructive proof of the theorem mentioned above and
it also contains different decomposition algorithms for multivariate
rational functions. In some sense, Problem \ref{prob-alg} can be
seen as a generalization of the univariate rational function
decomposition problem.

In this paper we will combine several techniques of Computational
Algebra to create an algorithm that finds all the intermediate
fields that are algebraic over the smaller field. Moreover, our
method can be extended to finitely generated $\K$-algebras, that is,
the case where the ambient field is $\K(z_1,\ldots,z_n)=\K(\z)$ for
some $z_1,\ldots,z_n$ transcendental over $\K$ that need not be
algebraically independent, and $\K(\z)$ is the quotient field of a
polynomial ring, so that we have
\[\K(\z)=QF\left(\K[x_1,\ldots,x_n]/I\right)\]
for some prime ideal $I\subset\K[x_1,\ldots,x_n]$ that will be given
explicitly by means of a finite system of generators.
Unsurprisingly, the algorithm will be much simpler when $\K(\x)$ is
rational, that is, when $I=(0)$.

\section{Main Results}

First, we can use Gr\"obner bases to compute and manipulate various
elements in our extensions, see \cite{Swe93} and \cite{BW93}. We can
compute transcendence and algebraic degrees of unirational fields,
decide whether an element is transcendental or algebraic over a
field, compute its minimum polynomial in the latter case, and decide
membership. Moreover, we  can compute  bases in the separable case
without using, properly, Gr\"obner bases, see \cite{Ste00}.

The next step is solving the problem when the given extension is
algebraic. We can rewrite the fields in the following way:

\begin{itemize}
    \item There exist rational functions $\hat\alpha_1,\ldots,\hat\alpha_n$
such that $\K(\hat\alpha_1,\ldots,\hat\alpha_n)/\K$ is a purely
transcendental extension, with
\[\K(\hat\alpha_1,\dots,\hat\alpha_n)\subset\K(f_1,\dots,f_m)\subset\K(x_1,\dots,x_n).\]

    \item There exist $\hat\alpha_{n+1},f$ algebraic over
$\K(\hat\alpha_1,\ldots,\hat\alpha_n)$ such that
\[\begin{array}{ll}
\K(f_1,\dots,f_m)= & \K(\hat\alpha_1,\ldots,\hat\alpha_n,\hat\alpha_{n+1}), \\
\K(x_1,\dots,x_n)= & \K(\hat\alpha_1,\dots,\hat\alpha_n,f).
\end{array}\]
\end{itemize}

Also, for any intermediate field in the extension there is $h$
algebraic over $\K(\hat\alpha_1,\ldots,\hat\alpha_n)$ such that
\[\F=\K(\hat\alpha_1,\ldots,\hat\alpha_n,h).\]

Thus, we can work in an algebraic simple extension. Let
$\E=\K(t_1,\ldots,t_n)$ be a purely transcendental field over $\K$,
$\E[\alpha]/\E$ an algebraic separable extension. Then, there exists
a bijection between the set of intermediate fields of
$\E\subset\E[\alpha]$ and the set of subgroups of the Galois group
$G$ that contain $G_\alpha$. Moreover, if
$\E[\beta],\E[\gamma]\subset\E[\alpha]$ are intermediate fields, we
can decide if $\E[\beta]\subset\E[\gamma]$.

It turns out that by factoring the minimal polynomial of $\alpha$
over $\E[\alpha]$, we can compute the intermediate fields of the
extension $\E[\alpha]/\E$. This is accomplished by means of using
decomposition blocks and, from the computational point of view,
factorization of polynomials in algebraic extensions, see
\cite{Tra76}, \cite{YNT89}, \cite{Rub01} and \cite{LM85}.

\begin{algorithm}\label{alg-algebraic} $ $

\noindent {\bf [A]} Factor $p_\alpha(z)$ in $E[\alpha]$.

\noindent {\bf [B.1]} If $p_\alpha(z)$ has more than one linear
factor:

$$p_\alpha(z)=(z-\alpha)(z-p_2(\alpha))\cdots(z-p_r(\alpha))
p_{r+1}(z,\alpha)\cdots p_{r'}(z,\alpha)$$

- Compute a minimal subgroup $G_{\psi}$ of
$<\{\sigma_2:\alpha\mapsto p_i(\alpha)\}>$.

- Consider $h(z)=\prod_{\sigma\in
G_{\psi}}(z-\sigma(\alpha))=a_ux^u+\cdots+a_0$.

- Take $a_i$ such that $\E[a_i]$ is a proper subfield of
$\E\subset\E[\alpha]$.

\noindent {\bf [B.2]}  If $p_\alpha(z)=(z-\alpha)p_2(z,\alpha)\cdots
p_{r'}(z,\alpha)$, with $p_i$ non-linear.

- Consider a factor  $P_2(z)=h(z, \alpha)(z-\alpha)$ of
$p_\alpha(z)$.

\[P_2=(z-\alpha)h(z,\alpha)=a_ux^u+\cdots+a_0.\]

- If $\E[a_i]= \E[\alpha]$ for all $i$, then  take another factor.
\end{algorithm}

In order to solve the general problem, we will compute the algebraic
closure of the given field in the ambient field. We will look for
the minimum field $\F_0$ that contains all the intermediate
algebraic fields over the given one. We adapt our data according to
the algorithm in  \cite{BV93} and \cite{Vas98}.

\begin{itemize}
    \item Let $h$ be the minimum common denominator of the rational
functions $f_i\in\K(\x)$.
    \item Let $\Phi:\ \K[y_1,\ldots,y_m]\to\K[x_1,\ldots,x_n,1/h]$,
defined as $\Phi(y_i)=f_i$ for each $i=1,\dots,m$.
    \item Let $\D_1=\Phi(\K[y_1,\ldots,y_m])=\K[f_1,\ldots,f_m]$.
We have that \linebreak $\D_1=\K[y_1,...,y_m]/\Ker(\Phi)$ is a
finitely generated $\K$-algebra. Also, the field of fractions of
$\D_1$ is $\K(f_1,...,f_m)$.
    \item Let $\D_2=\D_1[x_1,\ldots,x_n]=\K[x_1,\ldots,x_n,1/h]$.
The field of fractions of $\D_2$ is $\K(\x)$.
\item Let $t$ be a new variable and $\D=\D_1[t,x_1,\ldots,x_n]\subset\D_2[t]$, it
is a birational monomorphism.
 Compute the integral closure $\overline\D$ of the extension
$\D\subset\D_2[t]$ according to \cite{Vas98}. The integral closure
of the extension $\D_1\subset\D_2$ is $\D_0=\overline\D\cap\D_2$.
\item Then $\F_0$ is the field of fractions of $\D_0$.

\end{itemize}

Summarizing the results we have presented, we have the following
algorithm to find intermediate unirational fields over a given
field, if the extension is separable.

\begin{algorithm}\label{alg-general} $ $
\begin{description}
    \item \textsc{Input}: $f_1,\ldots,f_m\in\K(\x)$.
    \item \textsc{Output}: rational functions $h_1,\ldots,h_r$
such that
\[\K(f_1,\ldots,f_m)\varsubsetneq\K(h_1,\ldots,h_r)\varsubsetneq\K(\x).\]
\end{description}
\begin{description}
    \item \textsc{A}. Compute the algebraic closure of
$\K(f_1,\ldots,f_m)$ relative to $\K(\x)$.

    \item \textsc{B}. Find a separating basis of
$\K(f_1,\ldots,f_m)$.

    \item \textsc{C}. Rewrite the fields to obtain a simple algebraic
    extension.

    \item \textsc{D}. Factor the minimum polynomial obtained
in the algebraic extension.

    \item \textsc{E}. Compute the decomposition blocks that
correspond to the factors found before.

    \item \textsc{F}. If such a block exists, compute an intermediate field.

    \item \textsc{G}. Recover the generators of the intermediate
field in terms of the variables $\x$.
\end{description}
\end{algorithm}

Something that is worth mentioning is the fact that all the
computations can also be performed if the ambient field is not a
rational field but one of type $QF\left(\K[x_1,\ldots,x_n]/I\right)$
for some prime ideal $I$, the given extension being separable.
However, the theoretical and practical efficiency increases greatly,
since the representations of the elements are larger and all the
checks of type $f=0$ become $f\in\mathcal{B}_{\K(\x)/\K}$.

\section{Conclusions}

We have presented algorithms for resolving several issues related to
rational function field. Our approach has combined useful
computational  algebra tools. We also  unresolved many interesting
questions. Unfortunately, we do not know if the computed
intermediate field is rational or not, the reason is that the
algorithm produce an intermediate field generated always by the
transcendence degree plus one elements. Should be interesting to
investigate under which circumstances our algorithm can display an
intermediate subfield generated by as many elements as the
transcendence degree. From a more practical point of view, we would
like to have either a good algorithm or a good implementation to
compute a factorization of a polynomial over an algebraic extension.
Concerning applications, we regard the future interrelation of our
techniques to the factorization of morphisms and regular maps
between affine and projective algebraic sets.

\bibliographystyle{alpha}

\begin{vita}[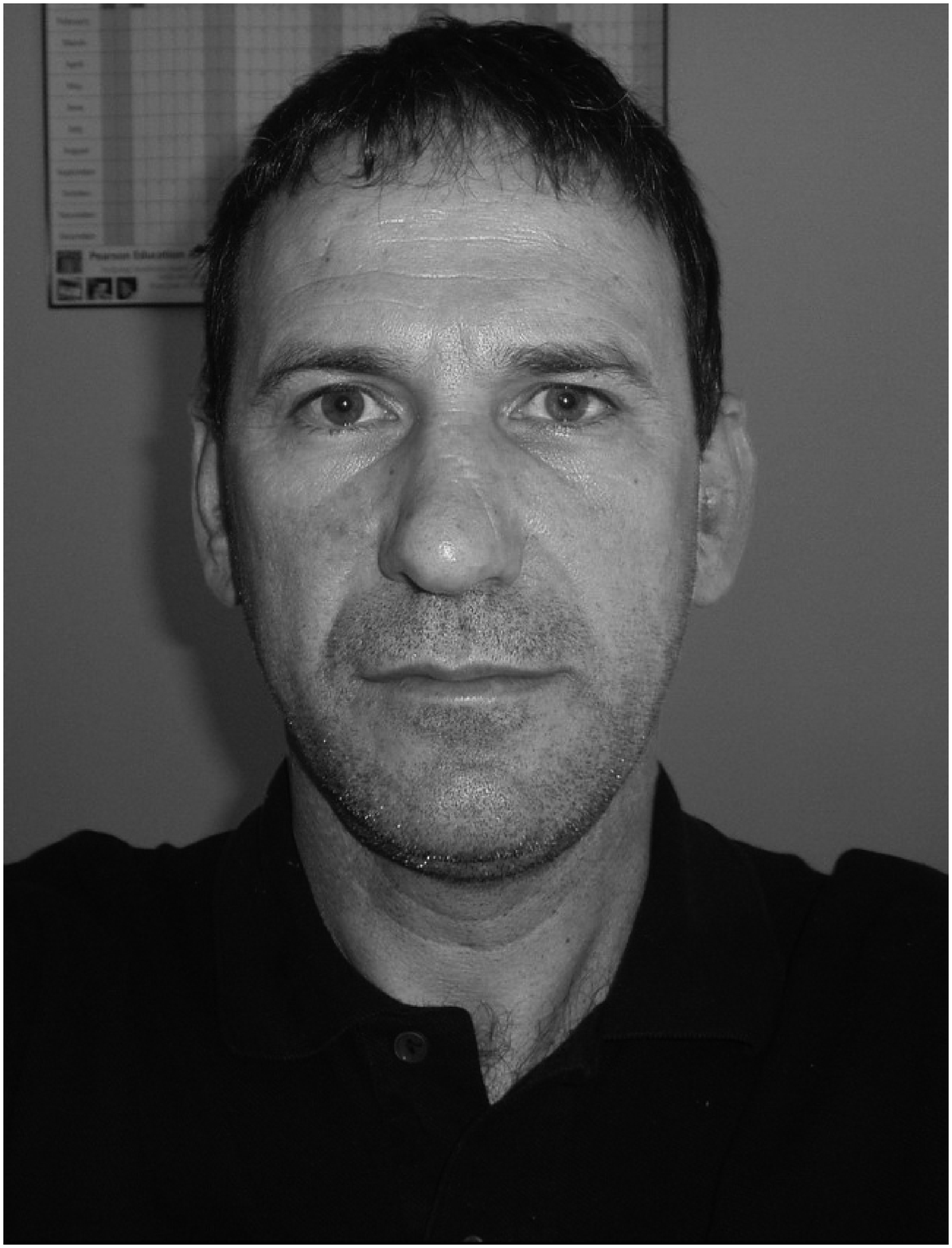]{Jaime Gutierrez}
{jaime.gutierrez@unican.es}{http://personales.unican.es/gutierrj/}
is an an Associate Professor of Mathematics at the University of
Cantabria, Spain, since 1991. His main interests are Computational
Algebra, Coding Theory and Cryptography.
\end{vita}
\begin{vita}[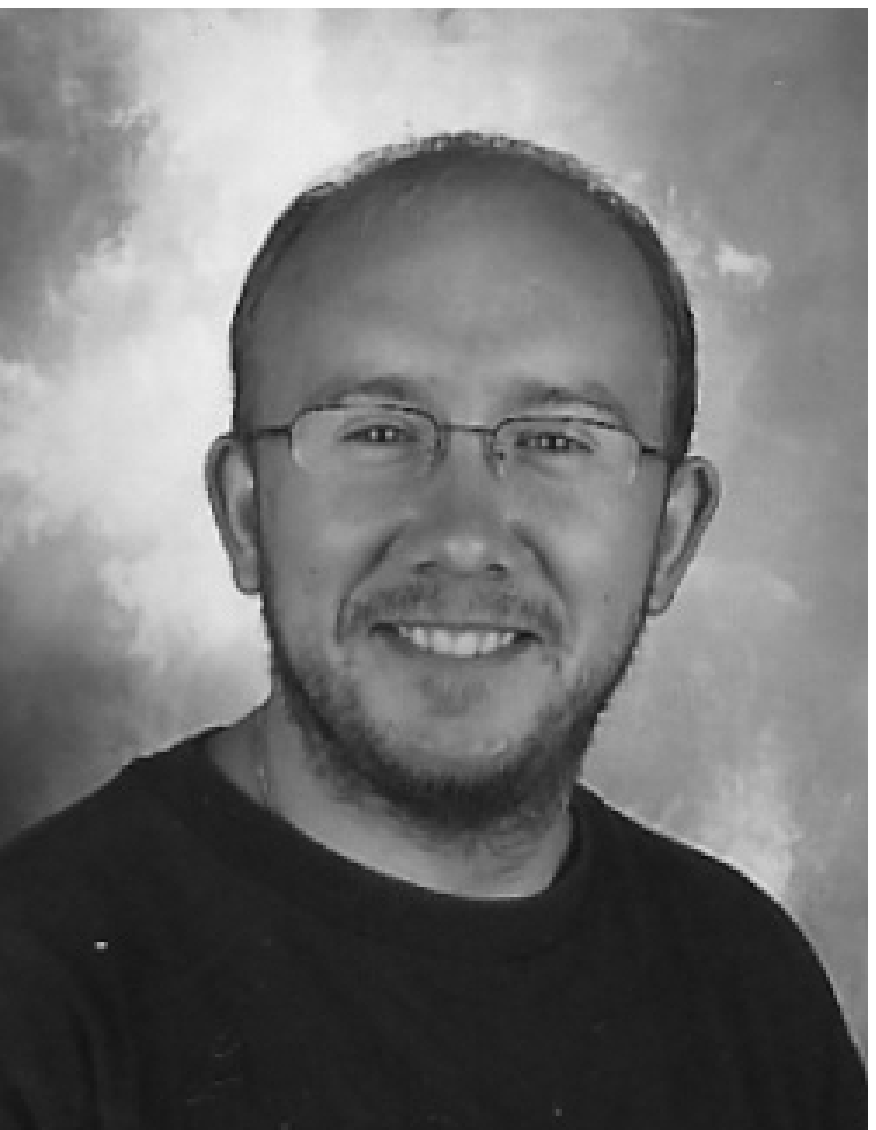]{David Sevilla}{david.sevilla@unican.es}{http://personales.unican.es/sevillad/}
is a Ph. D. in Mathematics since March 2004 and is currently
researching Functional Decomposition and other topics of
Computational Algebra in University of Cantabria, Spain.
\end{vita}
\end{document}